\def\C        {{$^{13}$C \/}}

\def\NN       {{$^{15}$N \/}}

\newcommand{\mr}[1]{\mathrm{#1}}
\newcommand{\unit}[1]{\,\mathrm{#1}}

\newcommand{\us}{\,\mu{\rm s}}
\newcommand{\uT}{\,\mu{\rm T}}

\newcommand{\ye}{\gamma_\mr{e}}

\newcommand{\ket}[1]{\ensuremath{\left|#1\right\rangle}}
\newcommand{\braket}[2]{\ensuremath{\left\langle#1|#2\right\rangle}}

\newcommand{\ketei}{\ket{0_\mr{e}}}
\newcommand{\keteii}{\ket{1_\mr{e}}}

\newcommand{\ketead}{\ket{\alpha'_\mr{e}}}
\newcommand{\ketep}{\ket{\psi_\mr{e}}}
\newcommand{\ketni}{\ket{0_\mr{n}}}
\newcommand{\ketnii}{\ket{1_\mr{n}}}
\newcommand{\ketna}{\ket{\alpha_\mr{n}}}
\newcommand{\RXpi}{R_X^\pi}

\newcommand{\cnote}{c-NOT$_\mr{e}$ \/}
\newcommand{\cnotn}{c-NOT$_\mr{n}$ \/}

\newcommand{\Tonee}{T_{1}}
\newcommand{\Tonen}{T_{1,\mr{n}}}

\newcommand{\Ttwoe}{T_{2}}
\newcommand{\Ttwon}{T_{2,\mr{n}}}
\newcommand{\apar}{a_\parallel}
\newcommand{\aperp}{a_\perp}
\newcommand{\wmwa}{\omega_\mr{mw,1}}
\newcommand{\wmwb}{\omega_\mr{mw,2}}
\newcommand{\wrfa}{\omega_\mr{rf,1}}
\newcommand{\wrfb}{\omega_\mr{rf,2}}
\newcommand{\ta}{t_\mr{A}}
\newcommand{\tb}{t_\mr{B}}
\newcommand{\tc}{t_\mr{C}}
\newcommand{\td}{t_\mr{D}}
\newcommand{\tsens}{t_\mr{meas}}
\newcommand{\fac}{f_\mr{ac}}

\newcommand{\Vpk}{V_0}

\newcommand{\captionstyle}{\normalfont} 

\documentclass[prl,aps,twocolumn]{revtex4-1}

\usepackage{graphicx}
\usepackage{bm}
\usepackage{amsmath}
\usepackage{amssymb}
\usepackage{verbatim}
\usepackage[colorlinks=true, pdfstartview=FitV, linkcolor=blue, citecolor=blue, urlcolor=blue]{hyperref} 
\usepackage{pifont}

\begin{document}

\global\emergencystretch = .1\hsize 

\title{A quantum spectrum analyzer enhanced by a nuclear spin memory}

\author{T. Rosskopf, J. Zopes, J. M. Boss, and C. L. Degen}
  \email{degenc@ethz.ch} 
  \affiliation{
   Department of Physics, ETH Zurich, Otto Stern Weg 1, 8093 Zurich, Switzerland.
	}
	
\date{\today}

\begin{abstract}
We realize a two-qubit sensor designed for achieving high spectral resolution in quantum sensing experiments.  Our sensor consists of an active ``sensing qubit'' and a long-lived ``memory qubit'', implemented by the electronic and the nitrogen-15 nuclear spins of a nitrogen-vacancy center in diamond, respectively.  Using state storage times of up to 45 ms, we demonstrate spectroscopy of external ac signals with a line width of 19 Hz ($\sim 2.9\unit{ppm}$) and of carbon-13 nuclear magnetic resonance (NMR) signals with a line width of 190 Hz ($\sim 74\unit{ppm}$).  This represents an up to 100-fold improvement in spectral resolution compared to measurements without nuclear memory.
\end{abstract}

\maketitle


Quantum sensors based on nitrogen-vacancy centers in diamond show promise for a number of fascinating applications in condensed matter physics, materials science and biology \cite{schirhagl14,rondin14}.  By embedding them in a variety of nanostructures, such as tips \cite{degen08apl,balasubramanian08,rondin12,maletinsky12}, nanocrystals \cite{dussaux16} or surface layers \cite{kolkowitz15,dussaux16}, local properties of samples can be investigated with high sensitivity and spatial resolution.  In particular, diamond chips with near-surface NV centers have enabled pioneering experiments in nanoscale detection of nuclear magnetic resonance (NMR), potentially enabling structural analysis of individual molecules with atomic resolution \cite{ajoy15,lazariev15,kost15}.

A key feature of many quantum sensing experiments is the ability to record time-dependent signals and to reconstruct their frequency spectra.  The canonical approach uses dynamical decoupling sequences, which are sensitive to frequencies commensurate with the pulse spacing while efficiently rejecting all other frequencies \cite{cywinski08,delange11,kotler11}. The spectral resolution of dynamical decoupling spectroscopy, however, is inherently limited by the inverse of the decoherence time $\Ttwoe$, which is about $(\pi\Ttwoe)^{-1} \sim 10-100\unit{kHz}$ for shallow NV centers \cite{loretz14apl}.
It has recently been recognized that by correlating two consecutive decoupling sequences, separated by a variable waiting time $t$, the spectral resolution can be extended to the inverse state life time $\Tonee$, which can be $10-100\times$ longer than $\Ttwoe$ (Refs. \onlinecite{laraoui11,laraoui13}).  Correlation spectroscopy has been applied to both generic ac magnetic fields and to nuclear spin detection, and spectral resolutions of a few 100 Hz have been demonstrated \cite{staudacher15,kong15,boss16}.

Despite these impressive advances there is a strong motivation to further extend the spectral resolution.  For example, many proposed nanoscale NMR experiments \cite{ajoy15,lazariev15,kost15} require discrimination of fine spectral features, often in the few-Hz range.  In addition, atomic-scale mapping of nuclear spin positions strongly relies on precise measurements of NMR frequencies and hyperfine coupling constants \cite{boss16}.  Therefore, methods to acquire frequency spectra with even higher spectral resolution are highly desirable.

In this study, we implement a two-qubit sensor designed to further refine the spectral resolution by a factor of $10-100\times$.  Our two-qubit sensor consists of an active sensing qubit and an auxiliary memory qubit, formed by the electronic spin and the \NN nuclear spin of the NV center in diamond.  By intermittently storing the state information in the nuclear -- rather than electronic -- spin qubit, we extend the maximum waiting time $t$ from $\Tonee \sim 1\unit{ms}$ to the nuclear $\Tonen > 50\unit{ms}$, with a corresponding gain in spectral resolution.  In addition, we use the nuclear memory to enhance sensor readout efficiency through repeated readout \cite{jiang09,lovchinsky16}, which would otherwise result in untenably long acquisition times.  The presented two-qubit system is particularly useful because it is intrinsic to the NV center, with no need for additional sensor engineering.

%
\begin{figure}[tb]
\includegraphics[width=1\columnwidth]{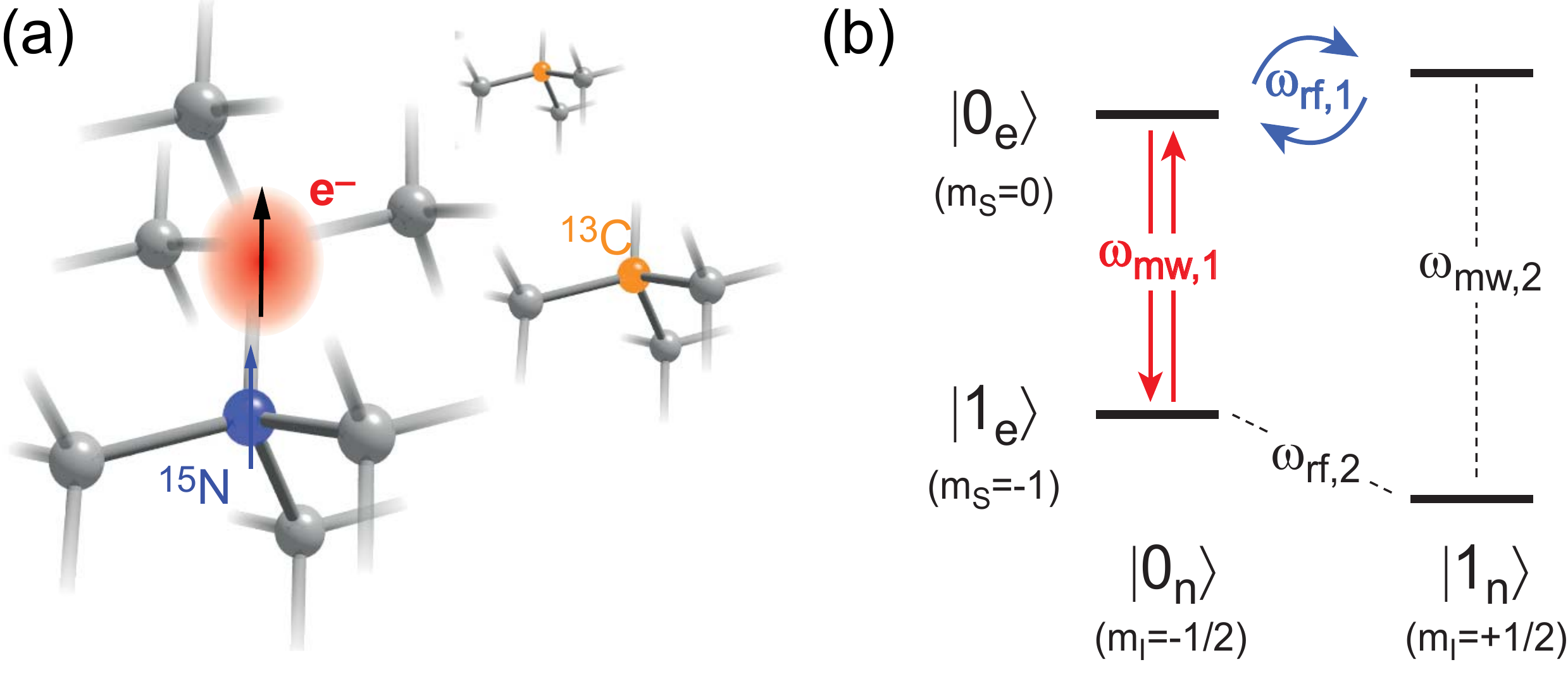}
\caption{\captionstyle
(a) Atomistic picture of the $^{15}$NV$^{-}$ two-spin system, showing the NV center's electronic spin (red) and \NN nuclear spin (blue).  Distant \C nuclei, which produce measurable NMR signals, are also shown.  Experiments are carried out on a single crystal diamond chip with a shallow ($3-10\unit{nm}$) layer of NV centers created by ion implantation.  
(b) Energy level diagram in the electronic ground state showing four resolved spin-flip transitions.  In a typical bias field of $320\unit{mT}$, aligned with the NV symmetry axis, the electron spin transition frequencies are $\wmwa=6097\unit{MHz}$ and $\wmwb=6100\unit{MHz}$, and the nuclear spin transition frequencies are $\wrfa=1.381\unit{MHz}$ and $\wrfb=1.669\unit{MHz}$, respectively.  Selective pulses use $\wmwa$ (red) and $\wrfa$ (blue).  Control pulses are applied via a coplanar waveguide connected to two separate arbitrary waveform generators \cite{supplemental}.
}
\label{fig:fig1}
\end{figure}
\begin{figure*}[tb]
\includegraphics[width=1\textwidth]{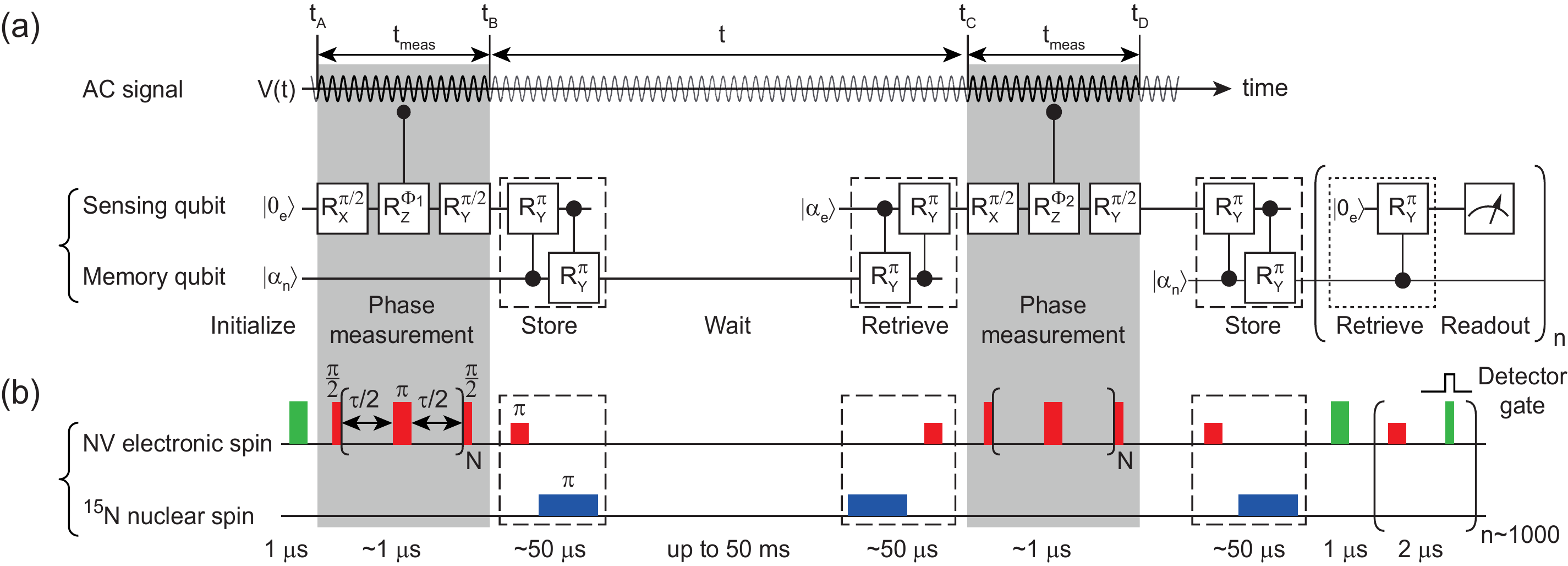}
\caption{\captionstyle
Implementation of the memory-assisted spectroscopy protocol for detecting alternating signals $V(t)$.
(a) Qubit gate diagram.  The core of the protocol are two phase gates $R_Z^{\Phi_1}$ and $R_Z^{\Phi_2}$ that are separated by a variable waiting time $t$, where $\Phi_1\propto V(\ta)$ and $\Phi_2\propto V(\tc)$.  The long-lived memory qubit allows extending the waiting $t$, which greatly enhances the Fourier-limited resolution of the spectroscopy protocol.  In addition, the memory qubit can be used to improve detection efficiency by a factor of $\sim n$ through repeated readouts. 
(b) Pulse timing diagram.  Laser pulses are shown in green, microwave pulses in red, radio-frequency pulses in blue, and the photon detector gate in a black contour.  The two phase measurements are implemented by XY8 sequences \cite{gullion90} with $N=8$ or $N=32$ pulses and $\tau\approx 1/(2\fac)$, where $\fac$ is the expected signal frequency.
}
\label{fig:fig2}
\end{figure*}
The advantage of one or more ``auxiliary'' qubits has been recognized in several recent works.  In particular, auxiliary nuclear spins have been used to increase the effective coherence time of an electronic sensor spin by quantum error correction \cite{unden16}, quantum feedback \cite{hirose16} or by exploiting double-quantum coherence \cite{zaiser16}.  Moreover, ancillary nuclei have been used to enhance the readout efficiency \cite{jiang09,lovchinsky16}.  In our study we utilize the auxiliary nuclear spin as a long-lived memory for the electron qubit's state.

Our two-qubit sensor exploits the four-level system formed by the $m_S \in \{0,-1\}$ subspace of the $S=1$ electronic spin and the two $m_I \in \{-1/2,+1/2\}$ states of the $I=1/2$ nuclear spin.  This pair has four allowed spin-flip transitions (see Fig. \ref{fig:fig1}).  Due to the hyperfine interaction ($\apar = 2\pi\times 3.05\unit{MHz}$), all four transitions are spectrally resolved and can be addressed individually using frequency-selective microwave or rf pulses.  Driving a selective $\RXpi$ rotation on either of theses transitions leads to a conditional inversion, depending on the state of the other spin.  This realizes controlled-NOT gates on the electronic and nuclear spins, respectively, which we denote by \cnote and \cnotn.

To implement the ``store'' and ``retrieve'' operations, we combine a \cnote gate and \cnotn gate (dashed boxes in Fig. \ref{fig:fig2}).  Assuming the electronic spin is initially in the $\ketei$ state and the nuclear spin in an idle (unspecified) state $\ketna$, the effect of the two gates is
$\ketei\ketna \xrightarrow{\text{\cnote}} \braket{0}{\alpha}_\mr{n} \keteii\ketni + \braket{1}{\alpha}_\mr{n} \ketei\ketnii$
$\xrightarrow{\text{\cnotn}} \braket{0}{\alpha}_\mr{n} \keteii\ketni + \braket{1}{\alpha}_\mr{n} \ketei\ketni = \ketead\ketni$,
where $\ketei\ketna$ etc. denote product states.
Likewise, if the electronic spin is initially in the $\keteii$ state, $\keteii\ketna \xrightarrow{\text{\cnotn\cnote}} \ketead\ketnii$.  As a result, the state of the electronic spin is stored in the state of the nuclear spin.  To retrieve the state, the order of the c-NOT gates simply needs to be reversed (see Fig. \ref{fig:fig2}).  Alternatively, the state can also be retrieved by initializing the electronic spin followed by a single \cnote gate (dotted box in Fig. \ref{fig:fig2}).
Opposite to the double c-NOT implementation, this protocol leaves the nuclear memory relatively unperturbed such that the memory state can be retrieved and read out many times \cite{jiang09,neumann10science}.

We assess the performance of the nuclear spin memory under a set of store, retrieve and hold operations.
To characterize the efficiency of the store and retrieve operations, we perform selective Rabi rotations on all four spin-flip transitions, and find efficiencies $>90\%$ for \cnote and $60-80\%$ for \cnotn, respectively \cite{supplemental}.
The memory access time is between $20-50\unit{\us}$ for the double c-NOT implementation, limited by the duration of the rf pulse, and $\sim 2\unit{\us}$ for the single c-NOT implementation.
We further test the complete memory by performing an electronic Rabi oscillation, storing the result in the nuclear memory, clearing the electronic qubit by an initialization step, and retrieving the Rabi signal \cite{supplemental}.  Lastly, we assess the memory hold time -- given by the nuclear $\Tonen$ -- in the absence and presence of laser illumination, with typical values of $\Tonen \approx 52\unit{ms}$ (no laser) and $\Tonen \approx 1.2 \unit{ms}$ (under periodic readout) at a bias field of $320 \unit{mT}$.  This bias field supports $n\sim 1000$ non-destructive read outs of the memory (dotted box in Fig. \ref{fig:fig2}) before the nuclear spin becomes repolarized \cite{supplemental,neumann10science}.

We compose the full spectroscopy protocol from a correlation sequence \cite{laraoui13,boss16} and several storage and retrieval operations (Fig. \ref{fig:fig2}).  In a first step, we initialize the electronic sensor spin into the $\ketei$ state.  An initial phase measurement is then performed using a multipulse sensing sequence approximately tuned to the frequency $\fac$ of the ac field (see Fig. \ref{fig:fig2}(b)).  During the multipulse sequence, the ac signal $V(t)=\Vpk\cos(2\pi\fac t)$ imprints a phase $\Phi_1 \propto V(\ta)$ on the electronic qubit, leaving it in a superposition $\ketep$ of states $\ketei$ and $\keteii$ with a probability amplitude $\braket{0}{\psi}_\mr{e}=\tfrac12(1+\sin\Phi_1)$.  Next, we store $\ketep$ in the nuclear memory, wait for a variable delay time $t$ (which can be very long), and read it back.  A second phase measurement is then used to acquire a further phase $\Phi_2 \propto V(\tc)$.  In a last step, we read out the final state of the electronic qubit via storing it in the nuclear memory and performing $n$ periodic readouts.  By averaging the protocol over many repetitions, the probability $p = |\braket{0}{\psi}_\mr{e}|^2$ of finding the sensor in the initial state $\ketei$ can be precisely estimated.

Because $\Phi_1$ and $\Phi_2$ depend on the relative phase of the ac signal $V(t)$, the total phase acquired by the qubit oscillates with $\fac$.  As detailed in the Supplemental Material \cite{supplemental}, the resulting state probability $p(t)$ then also oscillates with $\fac$,
\begin{align}
p(t)
  &= \frac12 \left( 1 - \langle \sin\Phi_1 \sin\Phi_2 \rangle \right)  \\
	&\approx \frac12 \left( 1 - p_0 \cos(2\pi\fac t) \right)  \label{eq:p}
\end{align}
where we assume that the ac signal is not synchronized with the acquisition.  Eq. (\ref{eq:p}) is for small signals where $\sin\Phi_1 \approx \Phi_1$ and $\sin\Phi_2 \approx \Phi_2$, resulting in an oscillation amplitude $p_0 \approx 2\ye^2\Vpk^2\tsens^2/\pi^2$ where $\tsens=\tb-\ta=\td-\tc$ (see Fig. \ref{fig:fig2}).  In order to obtain a frequency spectrum of $V(t)$, we can therefore simply measure $p(t)$ for a series of $t$ values followed by a Fourier transform.


%
\begin{figure}[t]
\includegraphics[width=1\columnwidth]{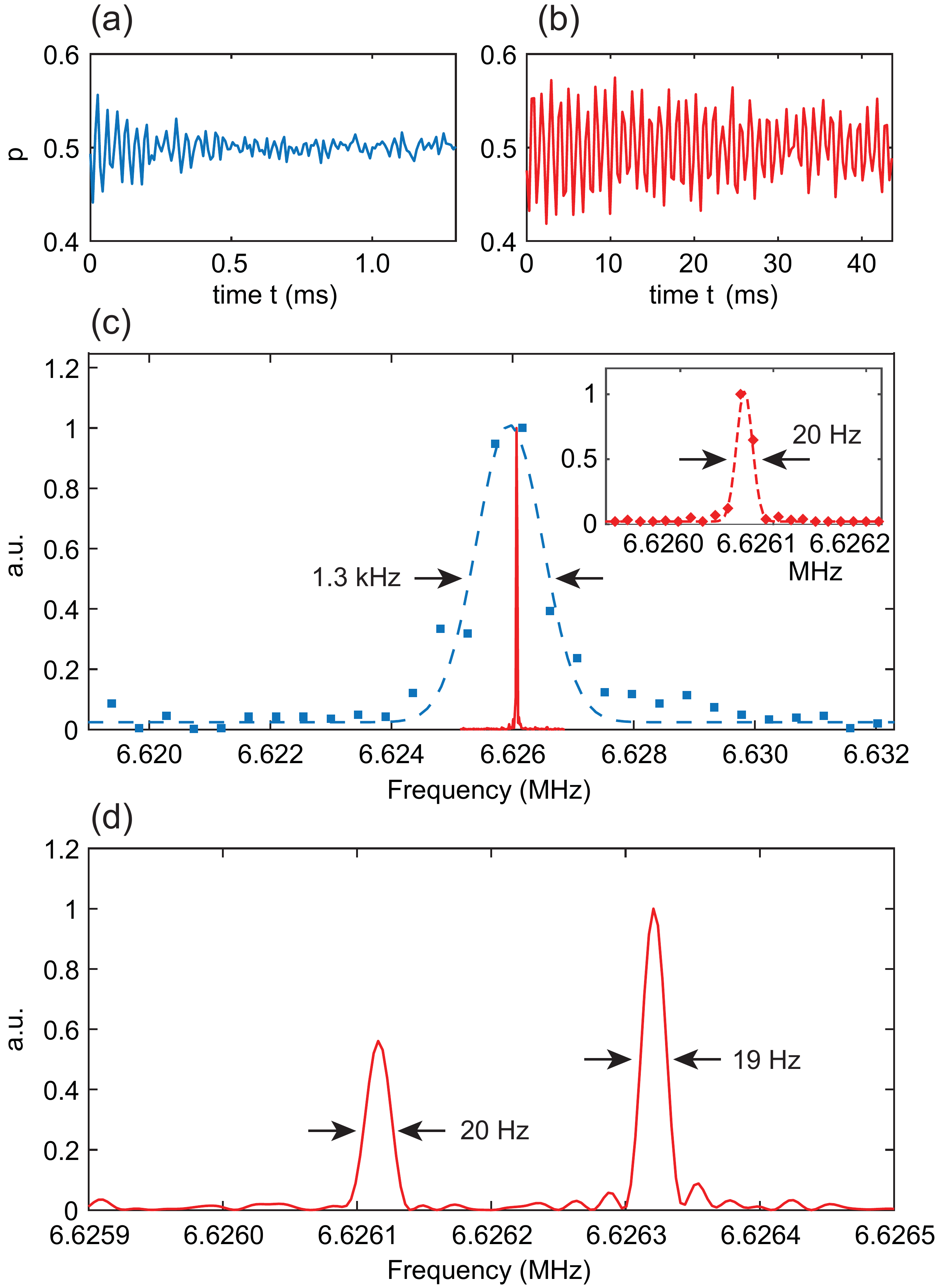}
\caption{\captionstyle
Frequency spectra of external ac test signals.
(a) Time trace recorded from an ac test signal without using the nuclear memory qubit.  A rapid decay of the signal is observed due to the electronic $\Tonee$ decay. 
(b) Time trace recorded from the same signal using the nuclear memory qubit.  The signal barely decays up to $t = 45\unit{ms}$.  Datapoints are sampled at a rate of $3.45\unit{kHz}$, corresponding to an undersampling by $1925\times$.
(c) Fourier transform (power spectrum) of the time traces from a) and b).  Inset shows a zoom-in for b).  Dots are the data and dashed lines are fits.  All line widths are full width at half height (FWHH). 
(d) Nuclear-memory-assisted spectrum of two artificial signals with fitted peak frequencies of $6.62611579(12)\unit{MHz}$ and $6.62632107(6)\unit{MHz}$.  The associated time trace is given in Ref. \onlinecite{supplemental}.  Total acquisition time per spectrum was on the order of 48 hours.
}
\label{fig:fig3}
\end{figure}			
We demonstrate the performance of the memory-enhanced spectrometer for two experimental scenarios.  In a first experiment, we expose the sensor to an external ac test signal with a nominal frequency of $\fac = 6.626070 \unit{MHz}$ and an amplitude of $\Vpk \approx 90\unit{\uT}$.  The test signal is produced on an auxiliary function generator not synchronized with the acquisition, and coupled into the same waveguide structure used for spin control.  Two measurements are carried out:  in a first acquisition (Fig. \ref{fig:fig3}(a)) we perform a regular spectroscopy measurement without the nuclear memory.  We can clearly observe an oscillation in the time trace due to the ac signal.  The signal decays on a time scale of $\sim 0.5 \unit{ms}$, limited by the electronic $\Tonee$ of this NV center.  In Fig. \ref{fig:fig3}(b) we repeat the measurement, now making use of the nuclear memory.  The oscillation persists beyond $t = 45 \unit{ms}$, overcoming the limitation due to the electronic $\Tonee$ by almost two orders of magnitude.

Fourier spectra of the two time signals (Fig. \ref{fig:fig3}(c)) show that the peak width reduces from $1.3\unit{kHz}$ ($200\unit{ppm}$) without memory to $20\unit{Hz}$ ($3.0\unit{ppm}$) with memory.  This corresponds to an improvement in spectral resolution by $\sim 65\times$.
Fig. \ref{fig:fig3}(d) shows a second example of nuclear-memory-assisted spectroscopy, where two ac test signals separated by about $0.2\unit{kHz}$ are applied.  Both peaks can be clearly distinguished, demonstrating that the method is effective in precisely resolving spectral features.  A narrow line width of only $19\unit{Hz}$ ($\sim 2.9\unit{ppm}$) is observed, and peak positions are defined with 8 digits of precision.  The absolute accuracy of the frequency measurement is governed by the internal clock of the microwave pulse generator.

The spectral resolution in Fig. \ref{fig:fig3}(c,d) is limited by the memory hold time given by the nuclear $\Tonen$, here $\sim 52\unit{ms}$.
Since the nuclear relaxation is dominated by a flip-flop process with the NV center's electron spin and slows down for higher bias fields \cite{neumann10science}, there is scope for an additional improvement in spectral resolution at Tesla bias fields \cite{pfender16}.


We further apply the two-qubit sensor to detect NMR spectra from nearby \C nuclear spins that are naturally present at $\sim1\%$ in the diamond chip.  This experiment represents an important test case towards the detection of more complex NMR spectra, such as those from molecules deposited on the chip \cite{mamin13,staudacher13,loretz14apl}.  The detection of nuclear spin signals is considerably more involved compared to external ac signals because the electronic sensor spin remains coupled to the nuclear spins during $t$ and causes back-action on the nuclear evolution.  In addition, NMR spectroscopy is known to be very sensitive to drifts in the external bias field.

\begin{figure}[t]
\includegraphics[width=1\columnwidth]{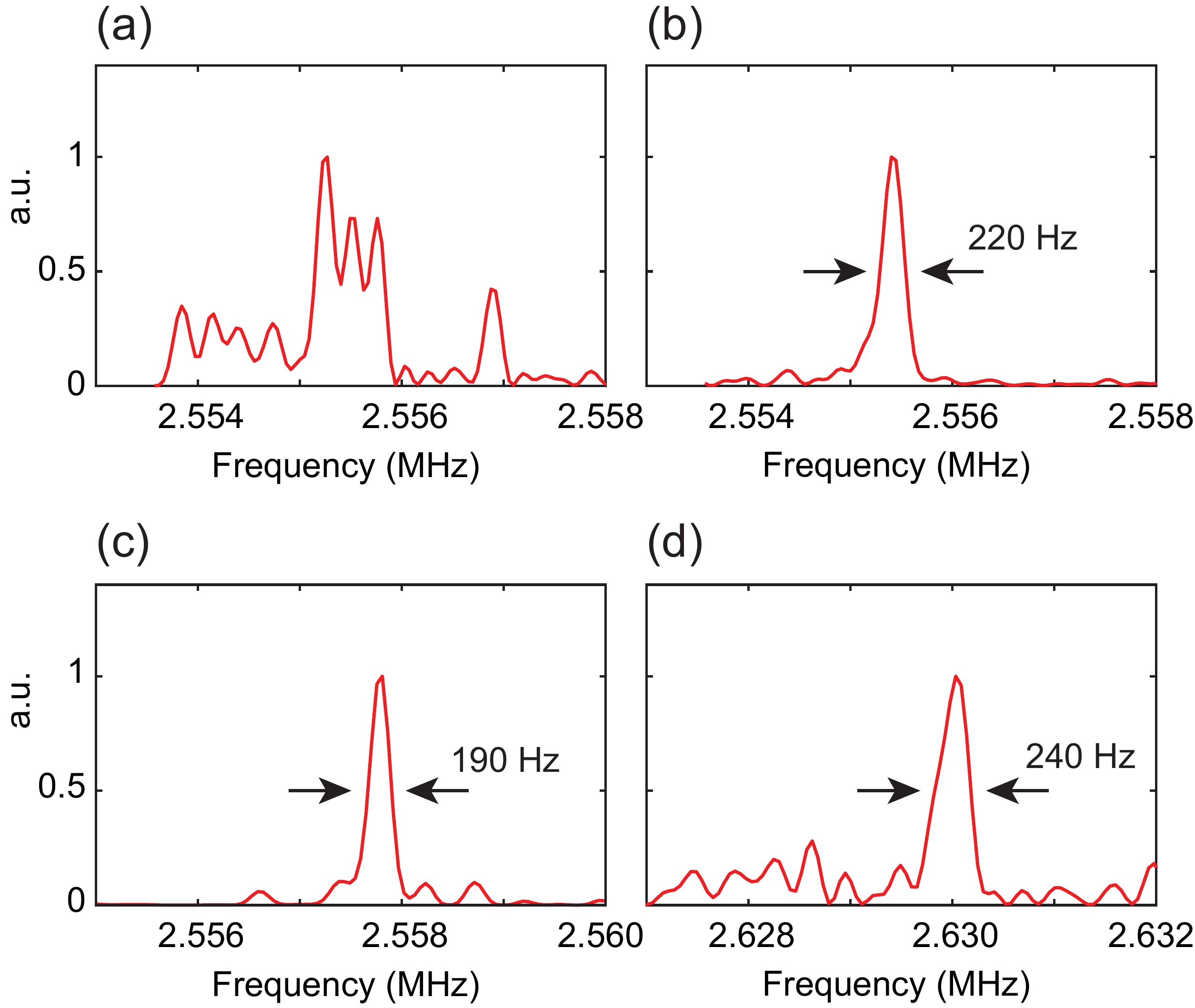}
\caption{\captionstyle 
NMR spectra (power spectra) of a nearby carbon-13 nuclear spin at a bias field of $240\unit{mT}$.
(a) Initial spectrum before correcting for NMR frequency drift.
(b) Same as a), after correcting for frequency drift.  $N=7$ dynamical decoupling pulses are applied to the $m_S=0 \leftrightarrow -1$ transition of the electronic sensor spin.
(c) Same as b), with a decoupling pulse applied every $4\unit{\us}$.  The linewidth of $190\unit{Hz}$ corresponds to $\sim 74\unit{ppm}$.
(d) Same as b), where the electronic sensor spin is periodically re-pumped into $m_S=0$ by $N=6$ laser pulses.  The maximum $t$ time was between $3-4 \unit{ms}$ and the electronic $\Tonee$ was $1.4 \unit{ms}$.  The hyperfine coupling parameters were $\apar = - 2\pi\times 138.9 \unit{kHz}$ and $\aperp = 2\pi\times 120.55 \unit{kHz}$, respectively.
}
\label{fig:fig4}
\end{figure}

Fig. \ref{fig:fig4} shows a set of four NMR spectra recorded from the same \C nuclear spin using the memory-assisted spectroscopy protocol.  The four panels represent increasing refinements in spectrum acquisition.  Fig. \ref{fig:fig4}(a) shows an initial \C spectrum that displays features over a wide frequency range of several kHz ($\sim 1000\unit{ppm}$).  We find these features to be linked to small changes in the bias field, probably caused by temperature-induced drifts in the magnetization of the permanent magnet in our setup.  By carefully tracking the electron spin resonance during the experiment and using post-correction, these drifts can be eliminated (Fig. \ref{fig:fig4}(b)) \cite{supplemental}.

The remaining line width of the \C resonance is on the order of $\sim 220\unit{Hz}$, which corresponds to $(\pi\Tonee)^{-1}$ of this NV center.  This reflects the fact that $\Ttwon\approx\Tonee$ due to the hyperfine interaction between the electronic and \C spins.  Thus, even if our spectrometer is technically capable of achieving a $10\times$ better spectral resolution, this improvement does not carry over to the \C spectrum.

To further reduce the line width of the \C resonance, we have explored several decoupling protocols, neither of which turned out to be effective.  A first protocol (Fig. \ref{fig:fig4}(b)) includes a series of $\pi$ pulses on the $m_S = 0 \leftrightarrow -1$ transition.  However, this protocol does not decouple the $m_S = 0 \leftrightarrow +1$ and $m_S = -1 \leftrightarrow +1$ transitions, and only a marginal improvement can be expected.  Indeed, no significant change is observed in the \C line width without and with dynamical decoupling even when using many hundred decoupling pulses (Fig. \ref{fig:fig4}(c)).  A more effective approach would be to simultaneously decouple both the $m_S = 0 \leftrightarrow -1$ and $m_S = 0 \leftrightarrow +1$ transitions using double-frequency irradiation \cite{mamin14}, but this control is not currently supported by our hardware.
Instead, we use a series of laser pulses to periodically repolarize the NV center into the $m_S = 0$ state (Fig. \ref{fig:fig4}(d)) or the $m_S=-1$ state (data not shown).  No narrowing of the \C resonance is observed with either protocol.  Hence, we speculate that the residual line width is either because decoupling is ineffective due to the relatively strong hyperfine coupling \cite{supplemental}, or due to interactions with other \C nuclei in the diamond chip.  The latter could be addressed by homonuclear decoupling sequences in future experiments \cite{slichter90,maurer12}. 


In summary, we implemented a two-qubit quantum sensor based on the electronic and \NN nuclear spins in diamond.  By operating the nuclear spin as a long-lived quantum memory, we achieve exceptionally high spectral resolution, with a best effort of $19\unit{Hz}$ or $2.9\unit{ppm}$.  The \NN spin forms a particularly suitable memory qubit because the nucleus is a natural part of the NV center, and because long storage times are possible combined with rapid memory access.  The ability to sense signals with high spectral resolution supports recent strides at detecting nuclear spin signals of nanoscale sample volumes, with possible applications in single-molecule NMR spectroscopy.



\vfill
The authors thank Kristian Cujia for fruitful discussions and Nicole Raatz, Sebastien Pezzagna and Jan Meijer for help with sample preparation.
This work was supported by Swiss NSF Project Grant $200021\_137520$, the NCCR QSIT, and the DIADEMS programme 611143 of the European Commission.

\noindent

\bibliography{C:/Christian/ETH/labview/library/library}






\end{document}


\large
\begin{center}

\textbf{{Supplementary Information for: \\ ``A quantum spectrum analyzer enhanced by a nuclear spin memory''}}

\normalsize

\vspace{5 mm}

T. Rosskopf, J. Zopes, J. M. Boss, and C. L. Degen

\textit{Department of Physics, ETH Zurich, Otto Stern Weg 1, 8093 Zurich, Switzerland}

\end{center}

\small



\section{Materials and Methods}

\subsection{Diamond chips}

Two different single-crystal-diamond chips are used in the study.  Sample A is an electronic-grade, natural abundance (1.1\% $^{13}$C) plate with a shallow NV layer created by $5\unit{keV}$ $^{15}$N$^+$ ion implantation and an $800\oC$ annealing step. 
The average NV center depth is $8\unit{nm}$ according to SRIM simulations.  Sample B is an electronic-grade, isotopically pure ($<$0.01\% $^{13}$C) plate with a similar NV layer as sample A.  To further reduce the NV-center-to-surface distance, sample B was subjected to an oxygen etch at $\sim 550 \oC$ (Ref. \onlinecite{loretz14apl}).  The average NV center depth was $3-5 \unit{nm}$. Both samples were baked at $465\oC$ in air before experiments to clean the surface.

\subsection{Measurement apparatus}

The measurement apparatus consists of a home-built confocal microscope with 532 nm laser excitation and 630-800 nm fluorescence detection using a single photon counter module.  The confocal microscope is equipped with a patterned, $50-100\unit{\um}$-wide coplanar waveguide transmission line for producing microwave and radio-frequency magnetic control pulses at the sample location.  Timing of experiments is controlled by the digital marker channels of a Tektronix AWG 5002 arbitrary waveform generator.  Photons are analyzed by time tagging arrival times and correlating them with the timing of the pulse sequence.  In addition, the microscope is equipped with a permanent NdFeB magnet to produce bias fields of up to $\sim 400\unit{mT}$.  The direction of the vector magnetic field is aligned via a mechanical xyz-stage.

To control the electronic and nuclear spin transitions of the NV center separate microwave (MW) and radio frequency (RF) channels are used, respectively.  Microwave pulses are generated on the arbitrary waveform generator at a $100\unit{MHz}$ carrier and upconverted to the desired GHz frequency using a local oscillator (Quicksyn FSW-0020) and a single-sideband mixer (Marki microwave IQ-1545).
For the RF pulses a National Instruments NI PCI 5421 arbitrary waveform generator is used, which directly synthesizes the desired pulses without a mixing step.  The two signals are amplified separately and then combined using a bias-T (Meca) before being connected to the microwave transmission line.  The output of the transmission line is terminated in a $50\unit{\Omega}$ load.  With this arrangement typical Rabi frequencies of $20-30 \unit{MHz}$ for the electron spin and $10-30\unit{kHz}$ for the \NN nuclear spin could be achieved.  We observed that the Rabi frequency of the \NN nuclear spin was significantly enhanced by the hyperfine interaction \cite{chen15}.

\subsection{Storage and retrieval operations}

\begin{figure}[h]
\includegraphics[width=0.9\columnwidth]{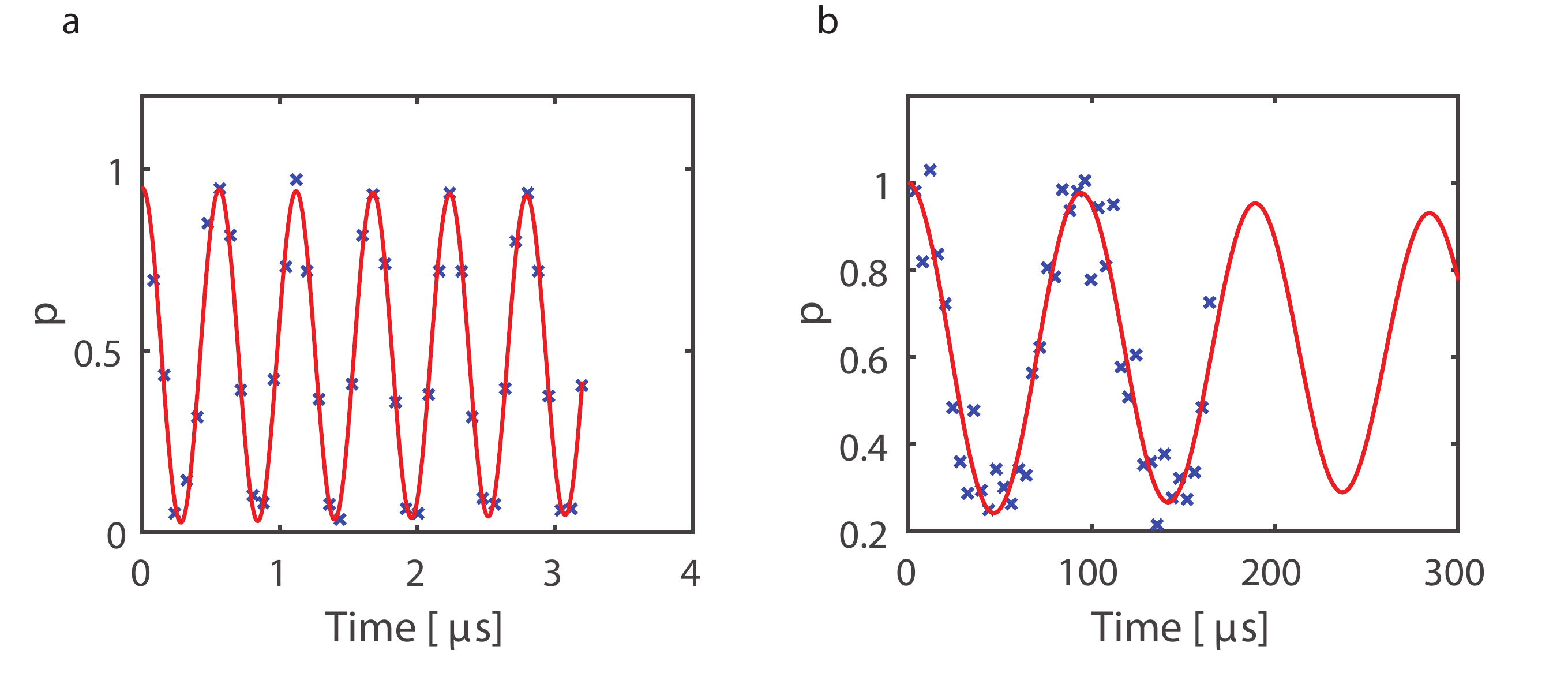}
\caption{\captionstyle Calibration of selective Rabi oscillations of (a) the electronic spin and (b) the nuclear spin.
The vertical scale is normalized to the maximum possible excursion from a separate calibration measurement.
For this NV center, the inversion efficiency was 92\% for the electronic spin and 77\% for the \NN nuclear spin, respectively.
Other NV centers had inversion efficiencies of $>$90\% for the electronic spin and 60-80\% for the \NN nuclear spin, respectively.
The efficiency for the electronic spin inversion is limited by spurious excitation of the other hyperfine resonance, while the efficiency of the nuclear spin inversion is limited by a combination of NV charge conversion \cite{aslam13} and imperfections of the rf pulse.
}
\label{fig:Suppfig5}
\end{figure}

Storage and retrieval operations are implemented by selective pulses on one of the two resolved hyperfine lines.  Square-shaped pulses are used for both electronic and nuclear spin manipulations.  Selective microwave pulses have a typical duration of $700\unit{ns}$ corresponding to a Rabi frequency of $\sim 0.7\unit{MHz}$.  Selective rf pulses have a duration between $20-50\unit{\us}$ depending on the set up and NV center.  A delay of $1-2 \unit{\us}$ is added after every radio-frequency pulse due to amplifier ringing.  Rabi oscillations used to calibrate the selective pulses are provided in Fig. \ref{fig:Suppfig5}.

\subsection{Correlation spectroscopy}

Correlation spectroscopy \cite{laraoui11,laraoui13,staudacher15,kong15,boss16} correlates the outcomes of two subsequent sensing periods to obtain high-resolution spectra of time-dependent signals.  In the present experiment, the method is implemented by subdividing a multipulse sequence into two equal periods of duration $\tmeas = \tb-\ta = \td-\tc$ that are separated by an incremented free evolution period $t$ (see Fig. 2 in the main manuscript).  Since the multipulse sequence is phase sensitive, constructive or destructive phase build-up occurs between the two sequences depending on whether the free evolution period $t$ is a half multiple or full multiple of the ac signal period $\Tac = 1/\fac$.  The final transition probability $p$  oscillates with $t$ as
\begin{align}
p(t)
  &= \frac{1}{2} \left( 1 - \langle \sin(\Phi_1)\sin(\Phi_2) \rangle \right)  \\
	&\approx \frac{1}{2} \left( 1 - \langle \Phi_1 \Phi_2 \rangle \right) \label{eq:papprox}
\end{align}
where
\begin{align}
\Phi_1 &= \frac{2\ye\tmeas}{\pi} V(\ta)  \nonumber \\
\Phi_2 &= \frac{2\ye\tmeas}{\pi} V(\tc)
\end{align}
Here, $V(t) = \Vpk \cos(2\pi\fac t + \phi)$ is an ac signal with amplitude $\Vpk$, frequency $\fac$ and phase $\phi$.  The second expression (\ref{eq:papprox}) is for weak signals where $\sin(\Phi_1) \approx \Phi_1$ and $\sin(\Phi_2) \approx \Phi_2$.  (For strong signals, higher harmonics appear in the correlation spectrum and the amplitude of the correlation signal does not directly reflect the amplitude of the ac signal any longer.)  The above expressions further assume that the multipulse sequence is tuned to the frequency of the ac signal, with an interpulse spacing of $\tau = 1/(2\fac)$.  An in-depth discussion of the technique can be found in, e.g., Ref. \onlinecite{laraoui13}.

In our experiments, the ac signal is not synchronized with the detection sequence, such that the phase $\phi$ is arbitrary.  Therefore, the observed correlation signal $\langle \sin(\Phi_1)\sin(\Phi_2) \rangle$ is an average over all $\phi = 0...2\pi$.  For the small signal approximation this leads to
\begin{align}
p(t)
  &\approx \frac{1}{2} \left( 1 - \langle \Phi_1 \Phi_2 \rangle \right)  \\
	&= \frac{1}{2} \left( 1 - \frac{4\ye^2\Vpk^2\tmeas^2}{\pi^2} \langle \cos(2\pi\fac\ta+\phi)\cos(2\pi\fac\tc+\phi) \rangle \right)  \\
	&= \frac{1}{2} \left( 1 - \frac{4\ye^2\Vpk^2\tmeas^2}{\pi^2} \langle \cos(\phi)\cos(2\pi\fac t+\phi) \rangle \right)  \\
	&= \frac{1}{2} \left( 1 - \frac{2\ye^2\Vpk^2\tmeas^2}{\pi^2} \cos(2\pi\fac t \right)  \\
	&= \frac{1}{2} \left( 1 - p_0 \cos(2\pi\fac t) \right) 
\end{align}
where we have set $\ta=0$ in the third step and used that $\cos(2\pi\fac\tc)=\cos(2\pi\fac[\tmeas+t])=\cos(2\pi\fac t)$.
Here, $p_0 = 2\ye^2\Vpk^2\tmeas^2/\pi^2$ is the amplitude of the correlation signal.

\subsection{Tracking and post-correction of spectral drift}

When performing NMR experiments we noticed considerable drifts, often several kHz, in the \C nuclear transition frequency.  Such frequency drifts are a well-known problem in high-resolution NMR spectroscopy, as they lead to unwanted line broadening.  Frequency drifts are typically caused by drifts in the static magnetic field; in our case, this is likely due to a temperature-related change of the magnetization of the permanent magnet.

We have implemented a tracking and post-correction scheme to eliminate the frequency drifts.  To follow the drift in magnetic field, we track the EPR resonance of the NV center during long-term measurements as shown in Fig. \ref{fig:FrequencyDrifts}.  The difference between measured EPR frequency $\we$ and reference EPR frequency $\weo$ corresponds to a drift in field by $\Delta B = (\we-\weo)/\ye$ where $\ye = 2\pi \times 28\unit{GHz/T}$ is the electron gyromagnetic ratio.  Measured datasets $p(t)$ are separately saved in intervals of $5-20\unit{min}$ with a specific $\Delta B$ tag for each dataset.

To correct for frequency drift, we multiply each dataset $p(t)$ by $e^{-i\Dw t}$, where $\Dw = \yn\Delta B$ is the expected shift in the \C nuclear resonance due to the drift in magnetic field.  $\yn = 2\pi \times 10.7\unit{MHz/T}$ is the \C gyromagnetic ratio and $T$ is the maximum $t$ time.  The corrected datasets $p(t)$ are then Fourier transformed and averaged.  Alternatively, the averaged NMR spectrum can also be obtained by performing a Fourier transform of the uncorrected $p(t)$ and shifting the frequency scale for each spectrum before averaging.

\newpage
\section{Supplemental Figures}

\subsection{Calibration of selective electronic and nuclear inversions}

\newpage
\subsection{Storage and retrieval of electronic Rabi oscillation}

\begin{figure}[h]
\includegraphics[width=0.50\columnwidth]{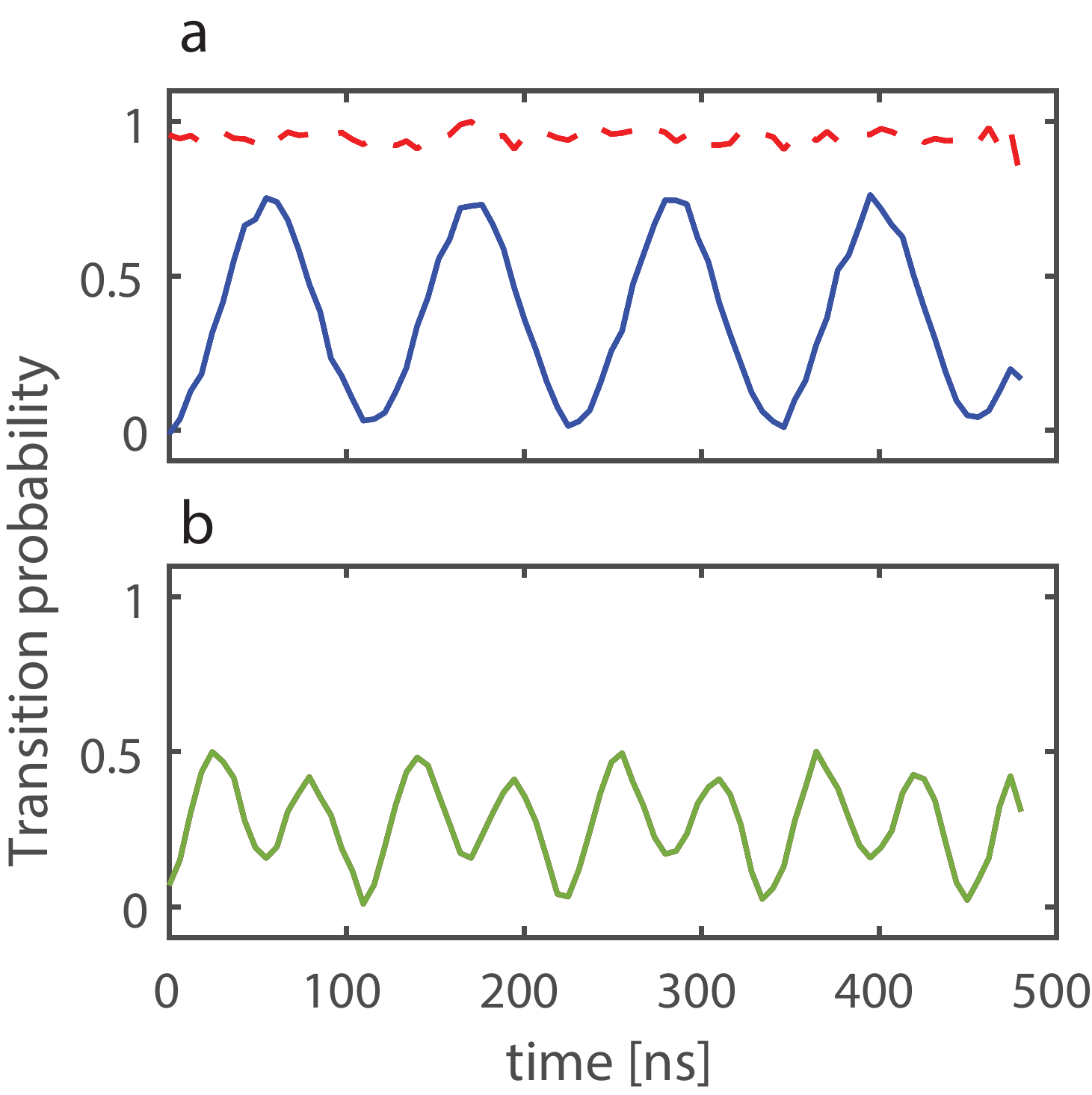}
\caption{\captionstyle
Test example of the nuclear quantum memory.
(a) Measurement to test the store and retrieve operations for the nuclear memory.  A Rabi oscillation is performed on the electron spin. The polarization is then transferred to the nuclear spin and the electron spin re-initialized by a laser pulse to clear.  The information is then retrieved and read out.  The blue curve shows the recovered signal.  The dotted red curve shows a control measurement without nuclear memory to verify that the re-initialization clears all information on the electronic spin.
The efficiency of the combined store and retrieve operation is only $\sim 70 \%$ in this case, because the laser pulse can induce conversion between the NV$^-$ and neutral NV$^0$ charge states \cite{aslam13}.  To avoid this effect, no laser pules were used during the main correlation sequence presented in Fig. 2 of the main manuscript.
(b) Measurement to test the correlation protocol.  After an initial Rabi rotation, the state of the electron is stored in the nuclear spin, the electronic spin re-initialized, and the state retrieved.  Then, a second identical Rabi rotation is performed before readout. An oscillation with twice the frequency is observed, as expected.  The apparent beating is due to the reduced efficiency caused by the laser pulse used for re-initialization.
}
\label{fig:Suppfig1}
\end{figure}	

\newpage
\subsection{Repetitive readout of nuclear spin state}

\begin{figure}[h]
\includegraphics[width=0.8\columnwidth]{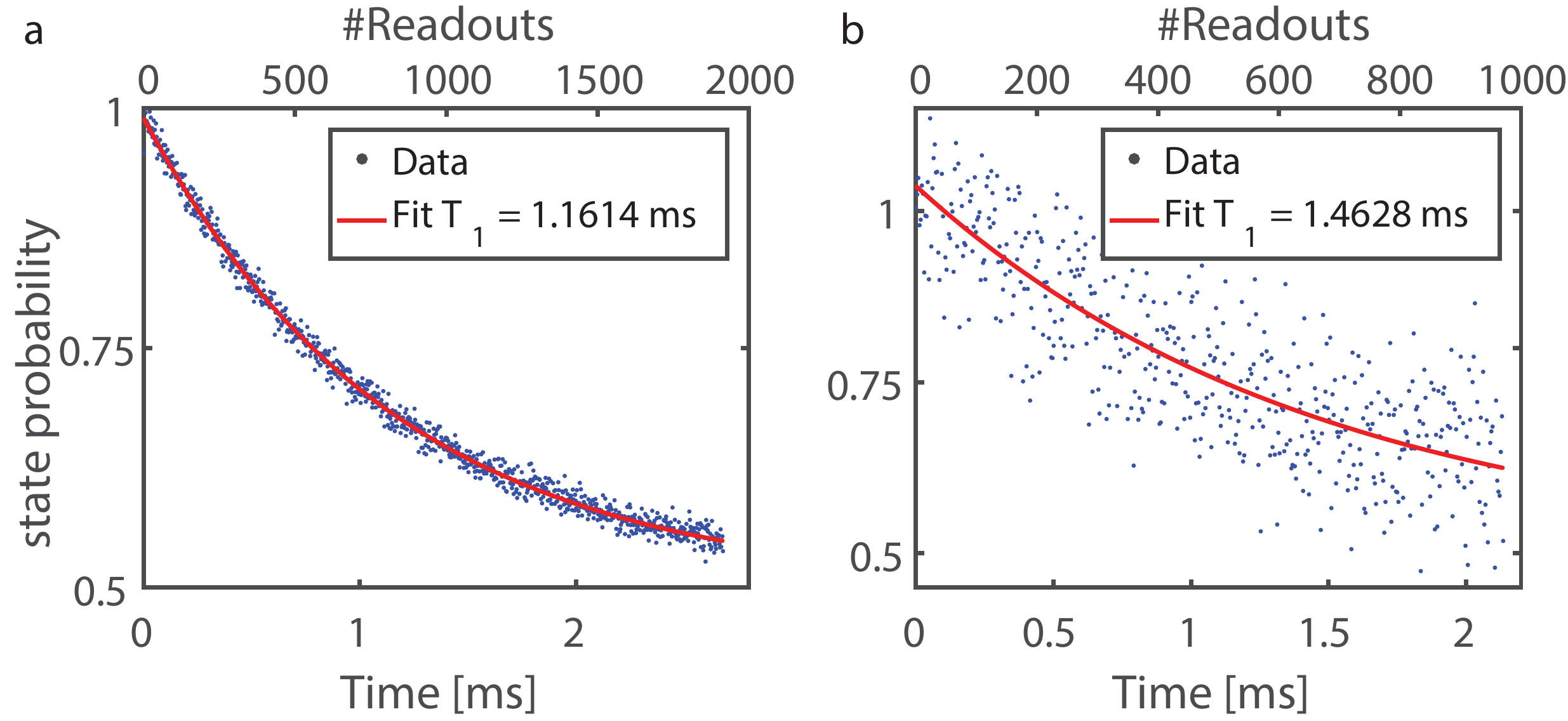}
\caption{\captionstyle
Repetitive readout of the nuclear spin state is performed to enhance the readout efficiency \cite{jiang09,neumann10science}.
Graphs show the state probability of the nuclear spin after a certain number of readouts $n$ are performed (see dotted box in Fig. 2 in the main manuscripts).  (a) is for a bias field of $\sim 350 \unit{mT}$ and (b) for a bias field of $\sim 250 \unit{mT}$.
}
\label{fig:Suppfig2}
\end{figure}

\newpage
\subsection{Additional ac sensing plots}

\begin{figure}[h!]
\includegraphics[width=0.9\columnwidth]{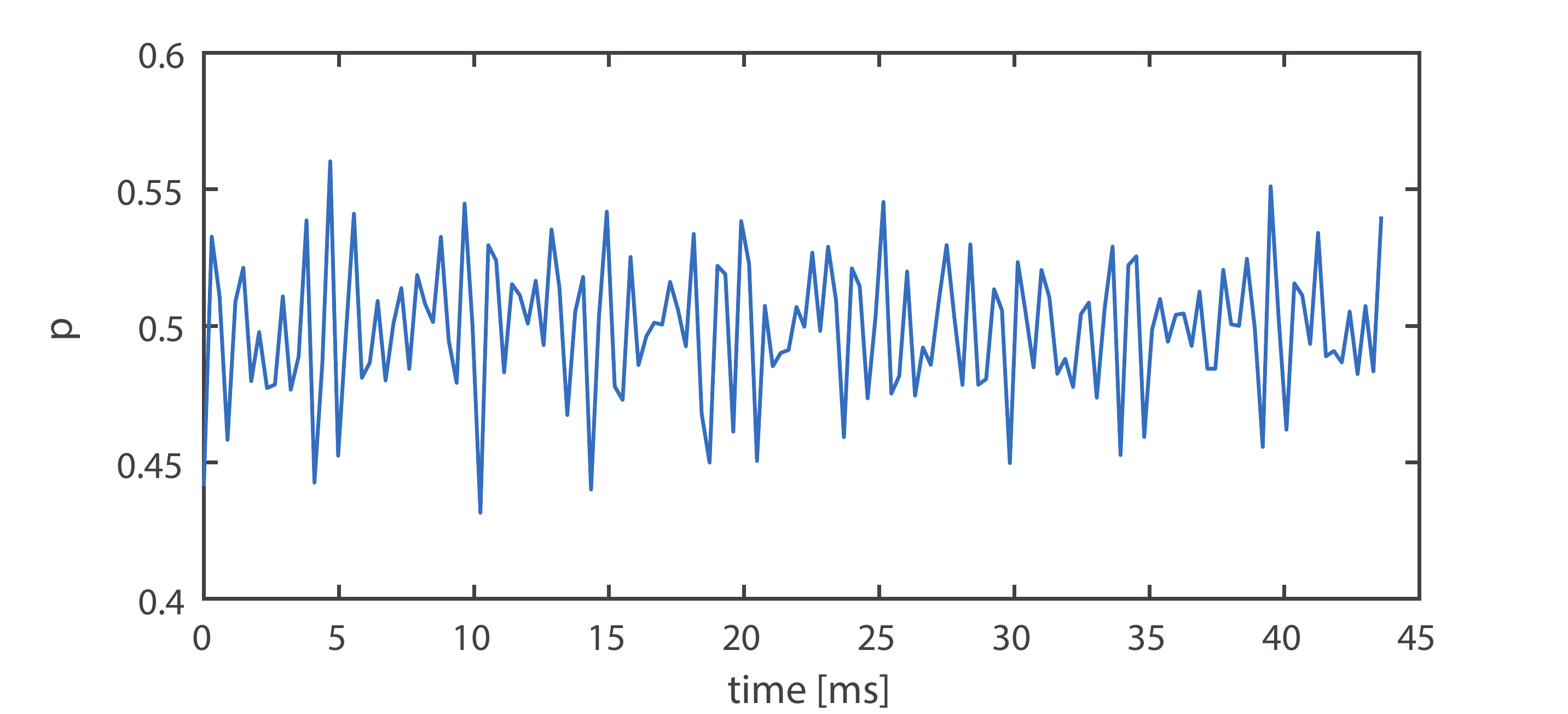}
\caption{\captionstyle Time trace for the spectrum presented in Fig. 3(d) in the main manuscript.  The beating due to the two frequency components can be nicely seen.
}
\label{fig:Suppfig4}
\end{figure}	

\newpage
\subsection{Tracking of magnetic field drifts}

\begin{figure}[htb]
\includegraphics[width=0.9\columnwidth]{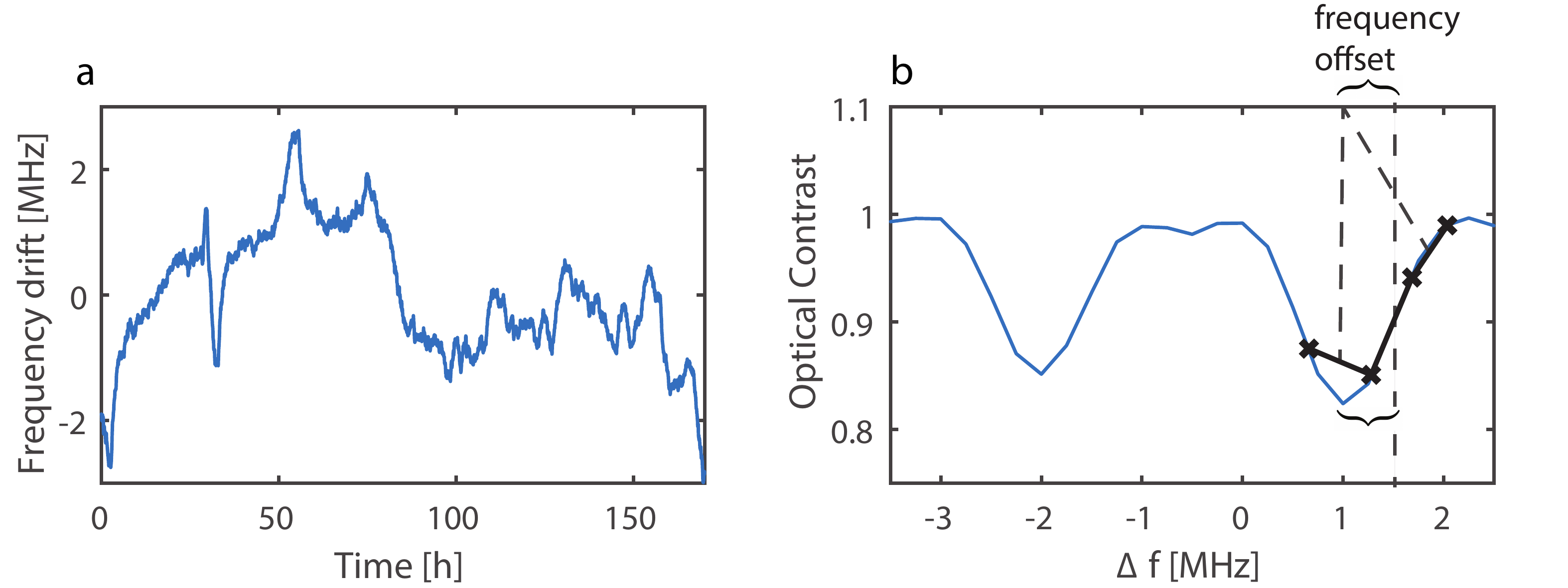}
\caption{(a) Relative frequency of the EPR resonance of the NV center over a long measurement period. 
The standard deviation of point-to-point differences is $33\unit{kHz}$, corresponding to an uncertainty in magnetic field of $1.2\unit{\uT}$ and an uncertainty in the \C resonance of about $13\unit{Hz}$.
(b) Protocol used to estimate the EPR frequency:  A four-point measurement of one hyperfine transition is performed to determine the peak position.  A simple tracking algorithm is implemented to follow the EPR peak as it shifts around due to magnetic field drifts.  The blue curve shows a representative EPR spectrum with both hyperfine lines, and the crosses indicate representative measurement points.
}
\label{fig:FrequencyDrifts}
\end{figure}

\newpage
\subsection{Simulation of \C NMR linewidth}


\begin{figure}[h]
\includegraphics[width=0.6\columnwidth]{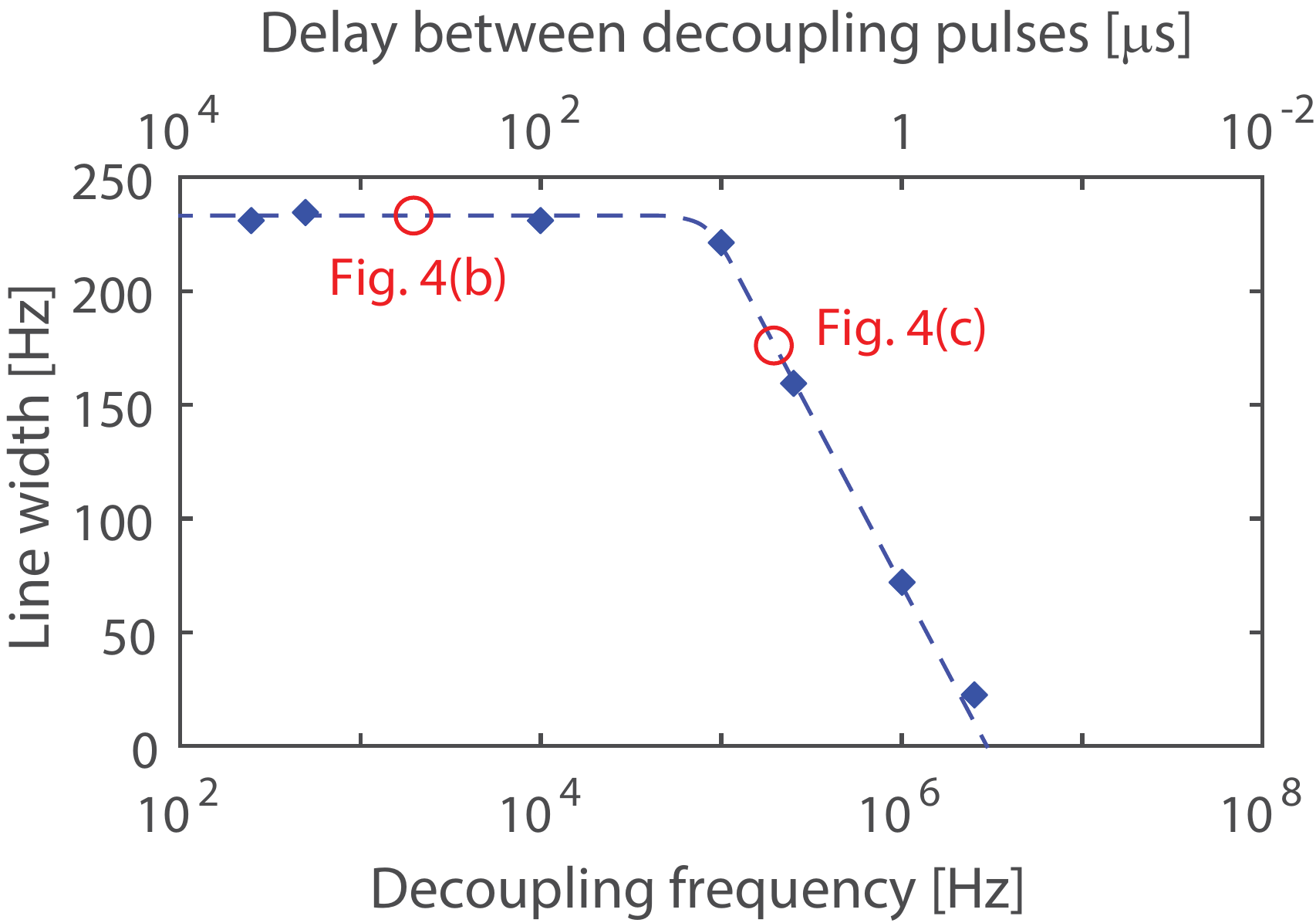}
\caption{\captionstyle
Simulation of the spectral line width of the \C NMR resonance with increasing number of decoupling pulses.
Only the $m_S=0\leftrightarrow-1$ transition is taken into account.
Magnetic field, hyperfine parameters and $\Tonee$ are set according to the experiment.
The simulation demonstrates that a line narrowing can be expected once the decoupling frequency, given by the inverse of the delay time between decoupling pulses, exceeds the hyperfine coupling (here $|\apar| \approx 2\pi\times 140 \unit{kHz}$, see main text).
Simulations use the density matrix method \cite{boss16}, $T_1$ decay is implemented by stochastic spin flips, and the obtained time traces are processed the same way as the experimental data.  Dots are fits to simulated spectra and the dashed line is a guide to the eye.  Red open circles indicate the approximate conditions for the spectra of Fig. 4(b) and (c) in the main manuscript, respectively.
}
\label{fig:Suppfig6}
\end{figure}


\newpage

\bibliography{C:/Christian/ETH/labview/library/library}